# Nanophotonic devices on thin buried oxide Silicon-On-Insulator substrates

**Suresh Sridaran and Sunil A. Bhave**

*OxideMEMS Lab, School of Electrical and Computer Engineering, Cornell University, Ithaca, NY 14850*
*ss625@cornell.edu*

**Abstract:** We demonstrate a silicon photonic platform using thin buried oxide silicon-on-insulator (SOI) substrates using localized substrate removal. We show high confinement silicon strip waveguides, micro-ring resonators and nanotapers using this technology. Propagation losses for the waveguides using the cutback method are 3.88 dB/cm for the quasi-TE mode and 5.06 dB/cm for the quasi-TM mode. Ring resonators with a loaded quality factor ($Q$) of 46,500 for the quasi-TM mode and intrinsic $Q$ of 148,000 for the quasi-TE mode have been obtained. This process will enable the integration of photonic structures with thin buried oxide SOI based electronics.

**OCIS codes:** (230.3120) Integrated optics devices (230.7370) Waveguides (250.5300) Photonic integrated circuits

## 1. Introduction

Silicon photonics has emerged as a promising solution for electronic photonic integration [1,2]. Several electronic photonic integrated devices have been demonstrated including GHz modulators [3-6], switches [7], detectors [8] and transceivers [6]. All demonstrations so far relied on large buried oxide to achieve optical confinement and low loss waveguiding [3-8]. However, commercial silicon-on-insulator (SOI), complementary metal oxide semiconductor (CMOS) electronics relies on thin buried oxide to reduce short channel effects [9] and avert thermal problems [10]. Hence, it is of great interest to demonstrate a photonic platform using standard thin buried oxide substrates to achieve intimate integration with front-end-of-line (FEOL) SOI CMOS electronics. In this paper, we demonstrate a high confinement photonic platform using thin buried oxide SOI wafers.

Traditionally, silicon strip waveguides are made on SOI with a thickness less than 260 nm and buried oxide thickness greater than or equal to 1 μm [11-13]. The waveguide width is defined lithographically and etched into silicon with a width less than 500 nm to ensure single-mode operation. The propagation loss reported for such waveguides is between 1-5 dB/cm [11,12] with values smaller than 1 dB/cm reported for waveguides made using oxidation smoothing [13]. The buried oxide on which the waveguide rests acts as a barrier preventing light leakage into the substrate and needs to be greater than 1μm to keep the substrate leakage loss to a value lower than 1 dB/cm [14,15]. For integration of waveguides on the same layer as SOI based CMOS, the thickness of buried oxide under the waveguides needs to match that used for electronics. The buried oxide for electronics is thinner and ranges from 100nm [16] to 400nm [17], which would give losses > 100dB/cm for the strip waveguides.

In this paper, we achieve optical waveguiding on a thin buried oxide SOI wafers using a localized substrate removal method. The substrate silicon is removed with an isotropic dry etch using xenon difluoride ($XeF_2$) gas [18].The approach followed is similar to that used to fabricate polysilicon waveguides above the trench isolation layer of a CMOS process [19,20]. We also report the fabrication of ring resonators with a loaded *Q* of 46,500 for the TM mode and an intrinsic *Q* of 148,000 for the TE mode.

## 2. Localized Substrate Removal for Nanophotonic Devices on Thin Buried Oxide SOI

The waveguides and ring resonators are fabricated on the 250 nm thick device layer of a SOITEC SOI substrate with a buried oxide thickness of 400 nm (Fig. 1). HSQ based negative e-beam resist (XR-1541) of thickness 130 nm is spin coated on top of the device layer. Waveguide are patterned on the resist using a JEOL 9300 e-beam lithography system and developed in AZ 300 MIF developer. The patterns are transferred to the silicon device layer using a $Cl_2$ based ICP etch. Photoresist is then spin coated to a thickness of 3 µm and substrate removal etch windows are patterned using a 5× i-line stepper. The etch windows are elliptical and are spaced 3 µm away from the waveguides. These windows are etched into the oxide using an RIE etcher with $CHF_3/O_2$ gases. The wafer is then diced before the substrate removal etch to allow for coupling through the ends of the waveguide. The substrate removal is performed in a XACTIX etcher by a pulsed xenon difluoride ($XeF_2$) etch recipe to obtain a lateral undercut of 15 µm.

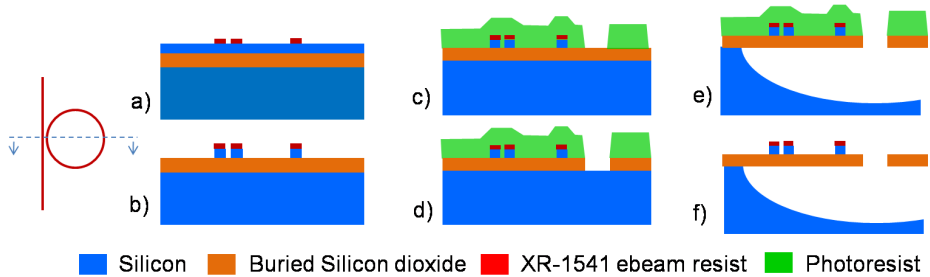

Fig 1. Process flow for waveguides and resonators on thin buried oxide SOI: a) Patterned e-beam resist, b) Transfer pattern into 250nm thick silicon device layer through $Cl_2$ ICP etch, c) Photoresist patterned to protect silicon device layer and open windows for localized substrate removal, d) Oxide etched using $CHF_3/O_2$ RIE to access substrate silicon, e) Cross-section after substrate removal through xenon difluoride etch, f) final device cross-section after photoresist strip.

The openings in the oxide act as windows for $XeF_2$ gas to reach the substrate. The undercut of silicon is decided depending on the desired air gap below the waveguide, the spacing between each of the etch windows and the spacing between etch windows and the waveguide. The photoresist protecting the device layer is removed after the $XeF_2$ etch by the use of oxygen plasma. Fig. 2 shows the SEM of a waveguide obtained with this process.

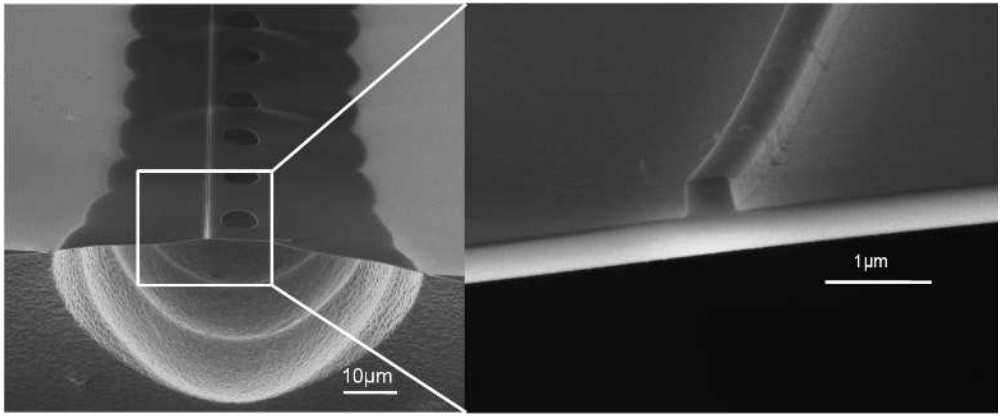

Fig 2. (left) SEM of waveguide showing etch windows in buried oxide and pits in substrate silicon created by $XeF_2$ etch, (right) Close-up of 250 nm silicon waveguide (with XR on top) resting on a 400 nm membrane of buried oxide.

## 3. Experimental Setup and Results

The transmission of light through the waveguides is measured for both quasi-TE and TM to characterize the waveguides and resonators. Light from a tunable laser is coupled into the waveguides via a polarization controller with the help of a tapered lensed fiber. The waveguides have a nanotaper [21] at the ends to convert the optical fiber mode into the waveguide mode. The light coming out of the waveguide is collected through a 40× microscope objective and focused on a detector via a polarizer. A loss of 18 dB is obtained from laser input to the detector for waveguides of 9 mm in length.

### 3.1 Waveguide Loss by Cutback Method

We obtain propagation losses using a cutback method [12] by fabricating meandering waveguides of different lengths. The cutback method is limited by the variations in coupling loss to different waveguides. Fig. 3 and Fig. 4 shows the data obtained from 2 different chips from the same wafer for the quasi-TE and TM modes. The loss obtained from the slope of the linear fit is 3.88 dB/cm for the quasi-TE and 5.06 dB/cm for the quasi-TM mode. This loss is comparable to silicon waveguides fabricated on buried oxide with thickness > 3 µm [12] showing that the loss due to substrate leakage has been eliminated in our process. The effective index obtained using COMSOL FEM software for the quasi-TE mode is 2.405 and the quasi-TM mode is 1.862 at $\lambda_0$=1550 nm.

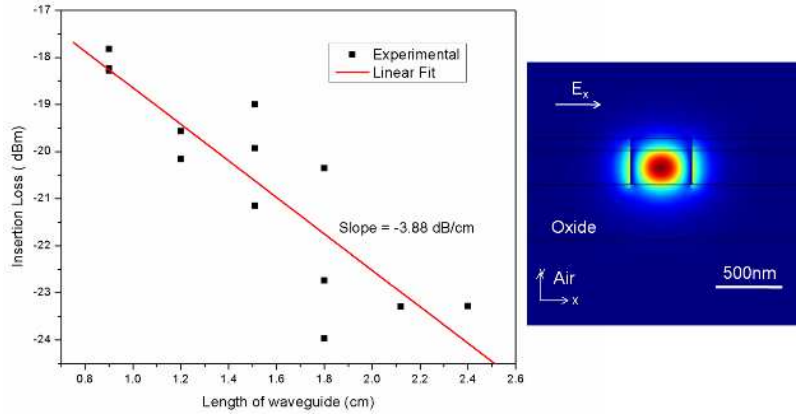

Fig 3. (left) Insertion loss for different waveguide lengths for quasi-TE mode. (right) Optical mode profile for the quasi-TE mode with effective index $n_{eff}$ = 2.405 calculated using COMSOL FEM software.

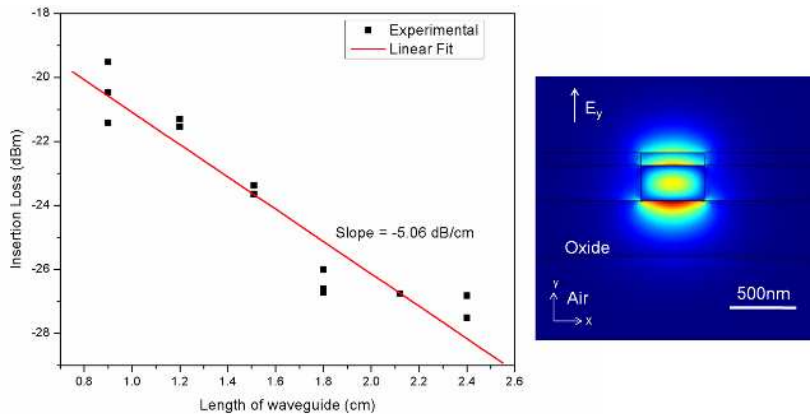

Fig 4. (left) Insertion loss for different waveguide lengths for quasi-TM mode. (right) Optical mode profile for the quasi-TM mode with effective index $n_{eff}$ = 1.862 calculated using COMSOL FEM software.

*3.2 Nanotapers*

The fiber mode is coupled into the waveguide mode by an inverted nanotaper as shown in Fig 5. The waveguide is tapered in a parabolic fashion from a size of 90nm at the tip to the waveguide width of 450nm over a distance of 40μm [21]. The loss for two of the nanotapers corresponds to the intercept of Fig. 3 and Fig. 4. The intercept for the TE mode is -14.77 dB corresponding to a loss of 7.385 dB per taper and for TM mode is -16.02 dB which gives a loss of 8.01 dB per taper. As seen from the SEM in Fig. 5, the nanotaper is etched 15 μm from the chip edge by the $XeF_2$ etch making the coupling lossy. This is because of insufficient photoresist coverage at the edge exposing the tip of the taper to the etching gases. This loss can be reduced in the future by slightly recessing the nanotaper from the edge of the chip thereby ensuring better protection by the photoresist.

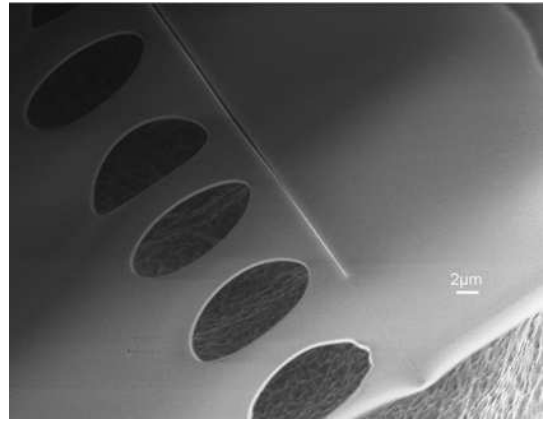

Fig 5. SEM of nanotaper for coupling from lensed fiber into waveguide.

*3.2 Ring Resonators*

The transmission spectrum of ring resonators was obtained by sweeping the tunable laser and recording the output of the waveguides from the detector onto a computer. The transmission spectrum for the quasi-TM mode for a ring resonator of radius 7.5 μm with a width of 450 nm and spaced 400 nm away from the waveguide is shown in Fig. 6. The free spectral range (FSR) for the resonances is used to calculate group index using $n_g = \lambda_0^2/(FSR \cdot L)$ [23], where $\lambda_0$ is the free space wavelength and L is the length of the resonator. The measured free spectral range is 10.44 nm from which the group index is calculated to be 4.92 around 1557.6 nm.

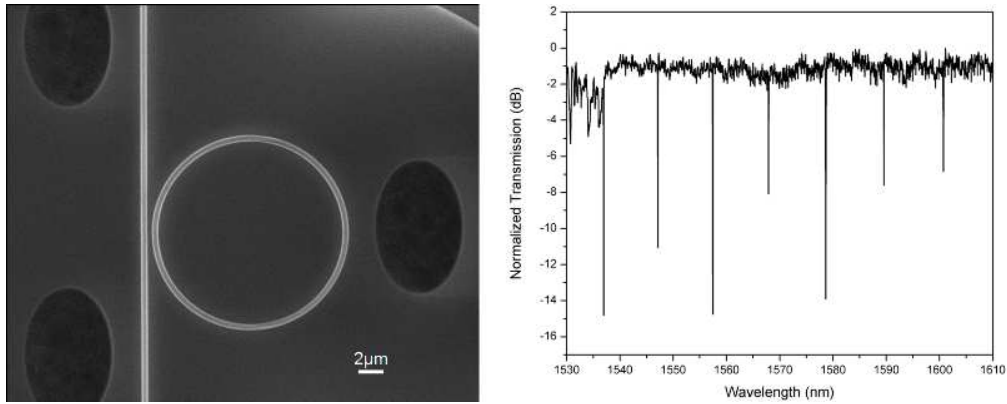

Fig 6. (left) SEM of ring resonator of 7.5μm radius spaced 400 nm away from the waveguide
(right) Normalized transmission spectrum of the ring resonator for the quasi-TM mode.

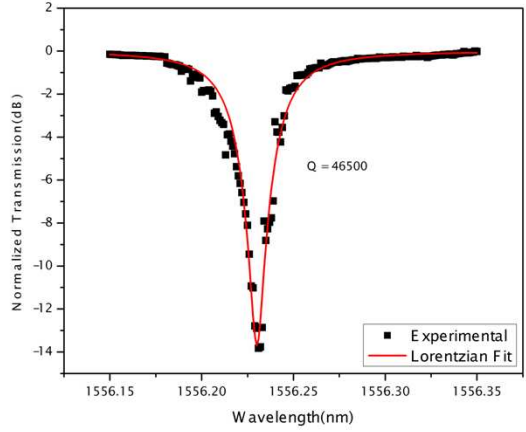

Fig 7. Normalized transmission for the quasi-TM resonance mode with a loaded Q of 46,500 and an extinction of 14 dB.

The loaded *Q* of 46,500 is obtained for the resonance shown in Fig. 7 by fitting a Lorentzian. At critical coupling the intrinsic *Q* is double the loaded *Q* giving an intrinsic *Q* of 93,000. Under critical coupling condition, the losses in the ring are calculated from the intrinsic quality factor using [22,23]

$$Q_{loaded} = \frac{Q_i}{2} = \frac{\pi n_g}{\lambda_o \alpha} \quad (1)$$

where α is the total propagation loss per unit length in the ring, and $n_g$ is the group index. The loss per unit length for the ring for a loaded *Q* of 46,500 is 9.28 dB/cm.

*3.3 Resonance splitting for quasi-TE mode*

The transmission spectrum for a quasi TE resonance of a 7.5 µm radius ring with a width of 450 nm and 150 nm spacing from the waveguide is shown in Fig. 8. A split resonance is obtained with a center wavelength of 1544.8 nm, a separation of 0.085 nm and an extinction ratio of 12 dB for each of the dips. Split resonances have been observed in ring resonators due to mode coupling between the clockwise and anticlockwise travelling modes [24]. The mode coupling is caused by surface roughness which is introduced due to imperfections on the resist and etching process. The transfer function for the transmitted power to the input power is given by [24, 25]

$$T(\omega) = \left|1 - \frac{1}{2Q_e}\left(\frac{1}{j\left(\delta + \frac{1}{2Q_u}\right) + \frac{1}{2Q_i} + \frac{1}{2Q_e}} + \frac{1}{j\left(\delta - \frac{1}{2Q_u}\right) + \frac{1}{2Q_i} + \frac{1}{2Q_e}}\right)\right|^2 \quad (2)$$

where $\delta = (\omega - \omega_0)/\omega_0$, $Q_i$ is intrinsic quality factor determined by loss per unit length in the ring, $Q_e$ is coupling quality factor determined by the rate of coupling of the mode to the waveguide and $Q_u$ is the mutual coupling quality factor which is controlled by the coupling between clockwise and anticlockwise propagating modes within the ring resonator. Eqn. (2) assumes that the two coupled resonances modes have the same resonance frequency and quality factor.

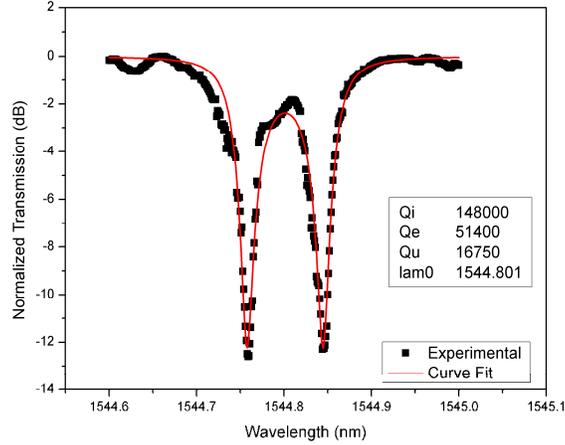

Fig 8. Normalized transmission for a quasi–TE resonance mode with the label describing the parameters obtained from fitting the measured data to Eqn. (2).

The free spectral range observed for the resonances is 11.15 nm and gives a group index of 4.54. The intrinsic quality factor obtained by fitting the measured transmission spectrum to Eq. (2) is 148,000. The propagation loss per unit length obtained from the intrinsic quality factor using Eq. (1) is 5.43 dB/cm.

## 4. Conclusion

We have demonstrated a photonic platform on thin buried oxide SOI wafers using localized substrate removal. Strip waveguides have been obtained with a loss of 3.88 dB/cm for the quasi-TE and 5.06 dB/cm for the quasi-TM mode. Nanotapers for coupling light in and out of the waveguides have been demonstrated. We have fabricated ring resonators with an intrinsic Q of 46,500 for the quasi-TM and 148,000 for the quasi-TE mode giving a loss of 9.28 dB/cm for quasi-TM and 5.43 dB/cm for the quasi-TE modes. The loss values measured for waveguides on the thin buried oxide are comparable to that obtained using thicker buried oxide. This gives the ability to make any nanophotonic device demonstrated on thick buried oxide SOI on a thin oxide CMOS SOI wafer, lowering the barriers to developing a common platform for CMOS electronics and Silicon photonics.

## 5. Acknowledgements

The authors wish to thank Sasikanth Manipatruni, Professor Michal Lipson and the Cornell Nanophotonics Group for useful discussions. This work was supported by the DARPA Young Faculty Award (Grant#: HR0011-08-1-0054) and was performed in part at the Cornell NanoScale Science and Technology Facility (CNF), a member of the National Nanotechnology Infrastructure Network (NNIN) (NSF Grant#: ECS-0335765).